\input phyzzx.tex
\tolerance=1000
\voffset=-0.0cm
\hoffset=0.7cm
\sequentialequations
\def\rl{\rightline}

\def\t1{{\tilde 1}}

\def\t{\theta}

\REF{\DSS}{G. Dvali, Q. Shafi and S. Solganik, hep-th/0105203.}
\REF{\BUR}{C. P. Burgess at. al. JHEP {\bf 0107} (2001) 047, hep-th/0105204; JHEP {\bf 0203} (2002) 052, hep-th/0111025.}
\REF{\ALE}{S. H. Alexander, Phys. Rev {\bf D65} (2002) 023507, hep-th/0105032.}
\REF{\EDI}{E. Halyo, hep-ph/0105341.}
\REF{\SHI}{G. Shiu and S.-H. H. Tye, Phys. Lett. {\bf B516} (2001) 421, hep-th/0106274.}
\REF{\KAL}{C. Herdeiro, S. Hirano and R. Kallosh, JHEP {\bf 0112} (2001) 027, hep-th/0110271.}
\REF{\SHA}{B. S. Kyae and Q. Shafi, Phys. Lett. {\bf B526} (2002) 379, hep-ph/0111101.}
\REF{\BEL}{J. Garcia-Bellido, R. Rabadan and F. Zamora, JHEP {\bf 01} (2002) 036, hep-th/0112147.}
\REF{\CAR}{K. Dasgupta, C. Herdeiro, S. Hirano and R. Kallosh, Phys. Rev. {\bf D65} (2002) 126002, hep-th/0203019.}
\REF{\JST}{N. T. Jones, H. Stoica and S. H. Tye, JHEP {\bf 0207} (2002) 051, hep-th/0203163.}
\REF{\LAS}{E. Halyo, hep-th/0307223.}
\REF{\BLO}{E. Halyo, hep-th/0312042.} 
\REF{\CON}{E. Halyo, hep-th/0402155.}
\REF{\BCQ}{C. P. Burgess, J. M. Cline, H. Stoica and F. Quevedo, hep-th/0403119.}
\REF{\QUE}{G. Aldazabal, L. E. Ibanez, F. Quevedo and A. M. Uranga, JHEP {\bf 0008} (2000) 002, hep-th/0005067; J. F. G. Cascales, M. P. Garcia del Moral, F. Quevedo and A. M. Uranga, 
hep-th/0312051.}
\REF{\PTE}{R. Kallosh, hep-th/0109168.}
\REF{\LIN}{R. Kallosh and A. Linde, JCAP {\bf 0310} (2003) 008, hep-th/0306058.}
\REF{\DTE}{E. Halyo, Phys. Lett. {\bf B387} (1996) 43, hep-ph/9606423.} 
\REF{\BIN}{P. Binetruy and G. Dvali, Phys. Lett. {\bf B450} (1996) 241, hep-ph/9606342.}
\REF{\TYP}{E. Halyo, Phys. Lett. {\bf B454} (1999) 223, hep-ph/9901302.}
\REF{\HAN}{A. Hanany and E. Witten, Nucl. Phys.{\bf B492} (1997) 152, hep-th/9611230.}
\REF{\KUT}{A. Giveon and D. Kutasov, Rev. Mod. Phys. {\bf 71} (1999) 983, hep-th/9802067.}
\REF{\FTE}{A. Linde and A. Riotto, Phys. Rev. {\bf D56} 1841 (1997); hep-ph/903209.}
\REF{\VAF}{F. Cachazo, S. Katz and C. Vafa, hep-th/0108120.}
\REF{\KAT}{S. Katz and D. Morrison, J. Algebraic Geometry {\bf 1} (1992) 449.}
\REF{\ASP}{P. S. Aspinwall, hep-t/9611137.}
\REF{\DOU}{M. Douglas, JHEP {\bf 9707} (1997) 004, hep-th/9612126.}
\REF{\DGM}{M. R. Douglas, B. R. Greene and D. R. Morrison, Nucl. Phys. {\bf B506} (1997) 84, hep-th/9704151.}
\REF{\TYE}{S. Sarangi and S. H. Tye, Phys. Lett. {\bf B536} (2002) 185, hep-th/204074; N. T. Jones, H. Stoica and S. H. Tye, Phys. Lett. {\bf B563} (2003) 6, hep-th/0303269;
G. Dvali and A. Vilenkin, hep-th/0312007; L. Leblond and S. H. Tye, hep-th/0402072.}
\REF{\POL}{E. J. Copeland, R. C. Myers and J. Polchinski, hep-th/0312067.}
\REF{\COS}{G. Dvali, R. Kallosh and A. Van Proeyen, hep-th/0312005.}
%\REF{\BRA}{E. Halyo, Phys. Lett. {\bf B461} (1999) 109, hep-ph/9905244; JHEP {\bf 9909} (1999) 012, hep-ph/9907223.}
%\REF{\JAU}{D. E. Diaconescu, M. R. Douglas and J. Gomis, JHEP {\bf 9802} (1998) 013, hep-th/9712230.} 
\REF{\STR}{E. Halyo, JHEP {\bf 0403} (2004) 047, hep-th/0312268.}
\REF{\MAT}{T. Matsuda, hep-ph/0402232; hep-ph/0403092.}
\REF{\PER}{M. Landriau and E. P. Shellard, astro-ph/0302166.}
\REF{\ACH}{J. Urrestilla, A. Achucarro and A. C. Davis, hep-th/0402032; A. Achucarro and T. Vachaspati, hep-th/9904229; A.Achucarro, A. C. Davis, M. Pickles and J. Urrestilla, hep-th/0109097.}
\REF{\DHK}{K. Dasgupta, J. Hsu, R. Kallosh, A. Linde and M. Zagerman, hep-th/0405247.}
\REF{\COM}{E. Halyo, in preparation.}
\REF{\NUT}{S. Hawking, Phys. Lett. {\bf 60A} (1977) 81; G. Gibbons and S. Hawking, Comm. Math. Phys. {\bf 66} (1979) 291.}
\REF{\QUI}{E Halyo, JHEP {\bf 0110} (2001) 025, hep-ph/0105216.}
%\REF{\STR}{A. Strominger, JHEP {\bf 010} (2001) 034, hep-th/0106113.}
%\REF{\INF}{A. Strominger, hep-th/0110087.}
%\REF{\HOL}{E. Halyo, hep-th/0203235.}

\singlespace
\rl{SU-ITP-04-}
\rl{hep-th/0405269}
\rl{\today}
%\rl{T}
\pagenumber=0
\normalspace
\medskip
\bigskip
\titlestyle{\bf{P--term Inflation on D--Branes}}
\smallskip
\author{ Edi Halyo{\footnote*{e--mail address: halyo@itp.stanford.edu}}}
\smallskip
\centerline {Department of Physics} 
\centerline{Stanford University} 
\centerline {Stanford, CA 94305}
%\centerline{and}
% \centerline{California Institute for Physics and Astrophysics}
%\centerline{366 Cambridge St.}
%\centerline{Palo Alto, CA 94306}
\smallskip
\vskip 2 cm
\titlestyle{\bf ABSTRACT}

We obtain a model of P--term inflation on D5 branes wrapped on resolved and deformed $A_n$ type singularities. On the brane world--volume, the resolution and deformation of the singularity 
correspond to an anomalous D--term and a linear term in the superpotential respectively. In the limiting cases with vanishing resolution or deformation we get F or D--term inflation
as expected. We give a T--dual description of the model in terms of intersecting branes.

\singlespace
\vskip 0.5cm
\endpage
\normalspace

\centerline{\bf 1. Introduction}
\medskip

There has been great interest in D--brane inflation[\DSS] in recent years mainly due to the possibility that we may be living on a D--brane. Recently
many models of D--brane inflation have been built[\DSS-\BCQ] which is a testament to the richness of this scenario. Moreover, it seems that D--brane inflation is the easiest way of 
realizing cosmological inflation in
string theory. It is therefore important to build generic D--term inflation models which can be realized in realistic string models (see [\QUE] for example). 

P--term inflation[\PTE,\LIN] is the generalization of D--term inflation[\DTE,\BIN,\TYP] to the case of ${\cal N}=2$ supersymmetry. The matter content of the model 
is given by a $U(1)$ vector multiplet and a
charged hypermultiplet. The superpotential and the Yukawa couplings are fixed by supersymmetry. In addition to the F and D--terms, the scalar potential can get a
contribution from a triplet of anomalous P--terms. In terms of ${\cal N}=1$ supersymmetry, one
of these can be seen as an anomalous D--term whereas the other two appear as a linear term in the superpotential. The model has a supersymmetric vacuum in addition to an unstable state
in which the neutral scalar describes a (classically) flat direction. In an inflationary scenario, the neutral scalar is the inflaton and its descent to the supersymmetric vacuum 
(due to the one--loop corrections to the scalar potential) describes inflation. In general both F and D-terms contribute to the scalar potential. However, for certain choices of 
parameters the linear term in the superpotential or the D--term vanishes and we obtain
D or F--term inflation respectively. An interesting property of P--term inflation is the fact that the F and D--terms can be mixed by a $U(2)$ transformation which exists due to ${\cal N}=2$
supersymmetry. 

We first show that P--term inflation can be obtained on D5 branes which live on spaces with $A_n$ (and possibly $D_n$ and $E_{6,7,8}$) type singularities. These compact spaces can be 
$ALE \times T^2$, an elliptically fibered Calabi--Yau manifold or a more complex space with a local $A_n$ singularity. As a concrete example, we consider the simplest case of an $A_2$
singularity, i.e. the smooth $Z_3$ ALE space ($\times T^2$). 
The smooth ALE space is obtained by Kahler and complex deformations of the $Z_3$ orbifold.
We show that the field theory on two D5 branes wrapping this resolved and deformed singularity gives rise to P--term inflation. This is a ${\cal N}=2$ supersymmetric theory with a $U(1) \times
U(1)$ gauge group and a charged hypermultiplet. The vector multiplets arise from strings that start and end on the same D5 branes wrapping each blown--up sphere. 
The hypermultiplets come from the strings
that connect the different D5 branes (which wrap different intersecting $P^1$'s). We show that (in ${\cal N}=1$ supersymmetric language) the origin of the anomalous D--term  
is the blow--up whereas the 
linear term in the superpotential arises from the complex deformation of the singularity. In this description, the transformation that mixes these two types of terms corresponds 
to the $SO(3)$ symmetry of the hyperKahler metric of the moduli space. 

We also obtain P--term inflation in terms of intersecting brane models[\HAN,\KUT] with two D4 branes stretched between three parallel NS5 branes. 
The two D4 branes correspond to the two D5 branes wrapping the $P^1$'s whereas the NS5 branes describe the smooth ALE space. The Kahler and complex deformations are now described by the 
positions of the NS5 branes along the three directions perpendicular to all branes. 
This description is related to the one in terms of wrapped branes by T--duality. 
Such a brane construction cannot be compactified and therefore describes only the physics near the resolved singularity. In this case, the transformation that mixes the F and D--terms is simply
a rotation, i.e. an $SO(3)$ transformation rotating the three directions transverse to all the branes.

The paper is organized as follows. In section 2 we review P--term inflation in ${\cal N}=2$ supersymmetric field theory. In section 3 we obtain P--term inflation on D5 branes which are 
wrapped on blown--up two cycles (of an orbifold type singularity) with complex
deformations. In Section 4 we describe P--term inflation in terms of intersecting brane constructions.
Section 5 contains our conclusions and a discussion of our results.

\bigskip
\centerline{\bf 2. P--term Inflation}
\medskip

P--term inflation[\PTE,\LIN] is the generalization of D--term inflation to ${\cal N}=2$ supersymmetry. The matter content of the model is given by a hypermultiplet and a gauge multiplet. These  
contain, in addition to the gauge boson, a pair of complex conjugate scalars $\Phi_A$, $(\Phi_A)^*$ and a neutral singlet scalar $\Phi_3$ respectively. The scalar potential 
including the ${\cal N}=2$ Fayat--Iliopoulos term is
$$V_P=2g^2[\Phi^{\dagger} \Phi \Phi_3^2+ {1 \over 4}(\Phi^{\dagger} \sigma_i \Phi -\xi_i)^2] \eqno(1)$$
where $\sigma_i$ are the Pauli matrices and $\xi_i$ are three anomalous P--terms. Renaming the scalars by $\Phi_1=\Phi_+$, $\Phi_2^*=\Phi_-$ and $S=\Phi_3$ and defining 
$\xi_{\pm}=\xi_1 \pm i\xi_2$, the scalar potential can be written in ${\cal N}=1$ supersymmetric notation as
$$V_P=2g^2(|S\Phi_+|^2+ |S \Phi_-|^2+ |\Phi_+ \Phi_- -{\xi_+ \over 2}|^2)+ {g^2 \over 2}(|\Phi_+|^2+|\Phi_-|^2- \xi_3)^2 \eqno(2)$$
The above potential can be written as a sum of an F--term and a D--term
$$V_P=|\partial W|^2+{g^2 \over 2} D^2 \eqno(3)$$
with the superpotential and D--term given by
$$W=\sqrt 2 g S (\Phi_+ \Phi_- -\xi_+/2) \qquad \qquad D=|\Phi_+|^2-|\Phi_-|^2-\xi_3 \eqno(4)$$
For $S>S_c=\xi/2$ the scalar potential has a nonsupersymmetric local minimum at
$$\Phi_+=\Phi_-=0 \qquad \qquad |P_i|^2=|(\Phi^{\dagger} \sigma_i \Phi -\xi_i)|^2=g^2\xi^2 \qquad \qquad V_0={1 \over 2} g^2 \xi^2 \eqno(5)$$ 
At this minimum, all supersymmetries are broken and therefore $V$ receives a one--loop contribution 
$$V_1={1 \over 2} g^2 \xi^2 \left(1+{g^2 \over {8 \pi^2}} log{|S|^2 \over |S_c|^2} \right) \eqno(6)$$
This one--loop correction to the scalar potential gives rise to a mass for the field
$S$. Thus, $S$ which plays the role of the inflaton, rolls--down its potential slowly resulting in slow--roll inflation. When $S<S_c$, the field $\Phi_+$ becomes tachyonic and starts rolling
towards its new minimum. The endpoint of inflation is the
supersymmetric minimum of the scalar potential with
$$S=0 \qquad \qquad |\Phi_+|^2={{\xi+\xi_3} \over 2} \qquad \qquad  |\Phi_-|^2={{\xi-\xi_3} \over 2} \eqno(7)$$

From the form of the scalar potential in eq. (2), it is clear that F--term[\FTE] and D--term[\DTE,\BIN,\TYP] inflation models are special cases of P--term inflation. 
From eqs. (2), (4) and (6) we see that
when $\xi_+=\xi_-=0$ we recover the D--term inflation scenario with the scalar potential
$$V_D=2g^2(|S\Phi_+|^2+ |S \Phi_-|^2+ |\Phi_+ \Phi_-|^2)+ {g^2 \over 2}(|\Phi_+|^2+|\Phi_-|^2- \xi_3)^2 \eqno(8)$$
where the Yukawa coupling is given by $\sqrt 2 g$ due to ${\cal N}=2$ supersymmetry.
On the other hand, the case with $\xi_3=0$ corresponds to the F--term inflation with the potential
$$V_F=2g^2(|S\Phi_+|^2+ |S \Phi_-|^2+ |\Phi_+ \Phi_- -M^2|^2)+ {g^2 \over 2}(|\Phi_+|^2+|\Phi_-|^2)^2 \eqno(9)$$
where we chose $\xi_+=\xi_-=M^2/2$.
Therefore, P--term inflation interpolates between F--term and D--term inflation models. 

An interesting property of P--term inflation is the $U(2)$ symmetry[\LIN] which arises from the underlying ${\cal N}=2$ supersymmetry. This can be used to show that F--term and D--term models
are related by a $U(2)$ transformation. Using eqs. (3) and (4) one can show that
$$V_F(\Phi)=V_D(\Phi^{\prime}) \eqno (10)$$
where
$$\Phi_3^{\prime}=\Phi_3 \qquad \qquad \Phi_A^{\prime}=U_A^B \Phi_B \qquad \qquad U={1\over \sqrt 2} (\sigma_1+\sigma_3) \eqno(11)$$ 

When the above model is coupled to ${\cal N}=1$ supergravity the scalar potential becomes[\LIN] (assuming canonical Kahler potentials for the fields)
$$\eqalignno {V=2g^2e^{|S|^2/M_P^2}[|\Phi_+ \Phi_- - \xi_+&/2|^2(1-(S \bar S/M_P^2)+(S \bar S/M_P^2)^2)\cr &+ |S \Phi_+|^2 + |S \Phi_-|^2]
+{g^2 \over 2}(|\Phi_+|^2-|\Phi_-|^2-\xi_3)^2 (12)}$$ 
For the inflationary trajectory in field space with $\Phi_+=\Phi_-=0$ including the one--loop correction to the scalar potential we get
$$V={{g^2 \xi^2} \over 2} \left(1+{g^2 \over {8 \pi^2}}log({|S|^2 \over |S_c|^2})+f({|S|^4 \over {2M_P^4}})+ \ldots \right) \eqno(13)$$
Coupling to gravity breaks the symmetry between the F and D--terms. The parameter 
$$f=(\xi_1^2+\xi_2^2)/\xi^2\eqno(14)$$ 
gives the relative strength of the F and D--terms in P--term inflation. We see that the cases with $f=0$ and $f=1$ correspond D--term and 
F--term inflation respectively.

\bigskip
\centerline{\bf 3. P--term Inflation on D--Branes}
\medskip

In this section, we obtain P--term inflation on D5 branes which are wrapped on a blown--up and deformed $A_2$ ALE singulaity. Consider the compact space $ALE \times T^2$
where the $Z_3$ ALE space in the orbifold limit is given by
$$f(x,y,z)=x^2+y^2+z^3=0 \eqno(15)$$
as a hypersurface in ${\bf C}^3$. This space is singular at $x=y=z=0$ which is the fixed point of the orbifold. There are two ways to remove this singularity. The first is by 
blowing up the singularity which means replacing the singular point by $P^1$'s ($S^2$'s). This is called a resolution or a Kahler deformation. The second is by deforming  
eq. (15) by adding a relevant deformation. This is called a complex deformation (or deformation for short) since it changes the complex structure of the space. The deformed form of eq. (15)
is[\VAF,\KAT,\ASP]
$$f(x,y,z,t_i)=x^2+y^2+\Pi_{i=1}^2(z+t_i)=0 \qquad \qquad t_1+t_2=0 \eqno(16)$$
For this new space, there is no solution to the equations $f=df=0$ and therefore the space is not singular. The $A_2$ singularity has only two such deformations which 
have to satisfy the
above cosntraint. The complex coordinates $t_i$ measure the ``holomorphic volume'' of the $P^1$'s in the geometry (which may or may not have a nonzero volume). It can be shown that 
the number of deformation coordinates
(two in our case) equals the number of $P^1$'s that can be blown up (also two). The ``holomorphic volume'' of the $P^1$'s is defined by[\VAF]
$$\alpha_i= \int_{P^1_i}{{dx dy} \over z} \eqno(17)$$
For each sphere this gives a complex number whose magnitude is the ``holomorphic volume''.

As mentioned above, we can also resolve the singularity. It is well--known that the number of 
$P^1$'s (which intersect each other pairwise) that are needed to completely resolve a singularity of type $A_n$ is $n$[\VAF,\ASP]. Thus we can resolve the singular space in eq. (15)
by blowing up two interecting $P^1$'s. The volume of each blown--up sphere is given by a real Kahler modulus 
$$v_i= \int_{P^1_i} K \eqno(18)$$
where $K$ is the Kahler form. The ``stringy'' volume of the resolved and deformed singularity is given by[\DOU]
$$V_i=(v_i^2+|\alpha_i|^2)^{1/2} \eqno(19)$$
where we assumed that $B_{NS}$ through the two spheres vanishes.

Now we consider two D5 branes, one wrapped on each of the two $P^1$'s of the above smoothed out singularity. The world--volume field theory is $3+1$ dimensional and has ${\cal N}=2$ 
supersymmetry. Since we have two separate D5 branes the gauge group is $U(1) \times U(1)$ with gauge couplings
$${1 \over g_i^2}={V_i \over {g_s \ell_s^2}} \eqno(20)$$ 
where $V_i$ is the ``stringy volume'' given in eq. (19). (There is no $\theta$ angle since we take $B_R$ flux to be zero.) In addition to the vector multiplets there is also a hypermultiplet
for each pair of intersecting $P^1$'s. We have exactly one intersection between the two $P^1$'s so we get one hypermultiplet in the bifundamental representation of the gauge group. 
In ${\cal N}=1$ supersymmetric terms there are two chiral multiplets with charges $(1,-1)$ and $(-1,1)$. Of the to $U(1)$'s,
the combination $1/2[U(1)_1+U(1)_2]$ describes the center of mass motion of the two D5 branes and decouples. The other combination given by $1/2[U(1)_1-U(1)_2]$ is the relevant one for
our purposes. Under this gauge symmetry, the two charge conjugate chiral multiplets ($\Phi_{1,2}$) have charges $1$ and $-1$. There is also a neutral chiral multiplet ($S$)
coming from the vector mulptiplet. Due to the ${\cal N}=2$ supersymmetry the superpotential is fixed to be
$$W=g_{YM} \Phi_1 S \Phi_2 \eqno(21)$$  
${\cal N}=2$ supersymmetry requires that the Yukawa coupling is given by the coupling of the $U(1)$ that does not decouple; $g_{YM}^{-2}=g_1^{-2}-g_2^{-2}$.

On the brane world--volume the Kahler and complex deformations of the singularity correspond to a triplet of anomalous P--terms as in eq. (1). Note that without these deformations
the ALE space is singular at $x=y=z=0$. This corresponds to the fact that $S=\Phi_+=\Phi_-=0$ is part of the moduli space. With the deformations, however, this singular point is
removed from the moduli space. Thus we expect the origin of the moduli space to be removed by modifications to the scalar potential. This can be achieved by adding a real anomalous 
D--term to the potential and a complex linear term to the superpotential. The moduli space of the resulting theory given by eq. (7) does not include the origin. 
In terms of ${\cal N}=1$ supersymmetry the deformation results in a linear term in the superpotential[\VAF]
$$W_1=\alpha_i S_i \eqno(22)$$
Again specializing to the combination $1/2[U(1)_1-U(1)_2]$ we get $W_1=\alpha S $
where $S=1/2[S_1+S_2]$ and $\alpha=[\alpha_1+\alpha_2]/2 \ell_s^4$. This together with eq. (21) gives exactly the superpotential of P--term inflation in eq. (4).
The resolution of the singularity corresponds to the anomalous D--term[\DGM]
$$\xi_3={1 \over {4 \pi^2 g_s}}{\sqrt v \over \ell_s^3} \eqno(23)$$
where $v=1/2[v_1+v_2]$. The total D--term becomes 
$$V_D=g^2(|\Phi_1|^2-|\Phi_2|^2-\xi_3)^2 \eqno(24)$$
exactly as in eq. (4).

We see that the scalar potential obtained from the above F and D--terms reproduces that of P--term inflation. 
Clearly if the singularity is only deformed (resolved) we get F--term (D--term) inflation. The moduli space of the world--volume field theory has a hyperKahler metric due to the ${\cal N}=2$
supersymmetry. Such a metric has an $SO(3)$ symmetry which rotates the three parameters $v$ and $\alpha$ into each other which is the symmetry in eq. (11).
Once coupled to gravity, the relative strengths of the F and D--terms is given by the parameter $f$ (see eq. (14))
$$f={{16 \pi^4 g_s^2 \ell_s^2 |\alpha|^2} \over {16 \pi^4 g_s^2 \ell_s^2 |\alpha|^2+v}} \eqno(25)$$
As expected $f=1$ ($f=0$) corresponds to F--term (D--term) inflation.

There are two main observational constraints on the string theory parameters of our P--term inflation model. These arise from the magnitude of density perturbations obtained from COBE and
the bound on the cosmic string contributions to this result. For P--term inflation we find that the Hubble constant during inflation is $H^2 \sim g^2 \xi^2/6M_P^2$ where 
$M_P^2=V_6/g_2^2 \ell_s^8$ ($V_6$ is the volume of the compact space.). The initial value of the inflaton
that will result in 60 e--foldings is
$$S_N= \xi+ {{g^2 N M_P^2} \over {2 \pi^2}} \eqno(26)$$
The COBE result on the magnitude of density perturbations
$$\delta_H={1 \over {5 \sqrt 3 \pi}} {V^{3/2} \over {V^{\prime}M_P^3}} \sim 2 \times 10^{-5} \eqno(27)$$
gives using eq. (12) for the scalar potential 
$${{2 \sqrt 2 \pi^2 \xi S_N} \over {g M_P^3}} \sim 5 \times 10^{-4} \eqno(28) $$ 
for $N=60$.

In the D--term inflation limit ($\alpha=0$ or $f=0$), from the form of the scalar potential it is clear that at the end of inflation the complex scalar field $\Phi_+$ can obtain any 
complex value with magnitude $xi_3$. Thus the vacuum manifold
is $S^1$ which leads to the formation of cosmic strings with tension $T= 2\pi \xi_3$ (for the physics of cosmic strings in D--brane inflation models see [\TYE,\POL]). These are not 
the recently discovered D--term strings[\COS,\STR,\MAT] even though they arise from D--terms since the superpotential of the model does not vanish. The superpotential can only vanish if the
compact space is not the direct product $ALE \times T^2$ but $ALE$ space fibered over $T^2$, such as a conifold[\STR]. In this case the fibration breaks supersymmetry to ${\cal N}=1$
and therefore we cannot have P--term inflation.
Cosmic strings generate density perturbations of the order of $O(G T)$. On the other hand, recent 
observations limit this contribution to at most $10^{-2}$ of the COBE result[\PER]. Thus we find $GT \sim M_P^2 T \leq 10^{-7}$ or $\xi_3 \leq 4 \times 10^{-7}M_P^2$. The constraint 
on string parameters is
$$\xi_3={1 \over {4 \pi^2 g_s}}{\sqrt v \over \ell_s^3} \leq 5 \times10^{-4} M_P^2 \eqno(29)$$
or
$${g_s^2 \over {4 \pi^2}}{{\ell_s^5 \sqrt v} \over V_6} \leq 10^{-7} \eqno(30)$$
We see that for $g_s \sim 1$ and $V_6 \sim \ell_s^6$ we find a very small blow up radius $\sqrt v \sim 10^{-5} \ell_s$ whereas for larger compactification radii, e.g. $V_6 \sim 10^6 \ell_s^6$
we get $\sqrt v \sim 4 \ell_s$ which is as large as the compactification radii.
Note that for $\sqrt v \sim 10^{-5} \ell_s$ the gauge coupling is extremely large, $g^2 \sim 10^{10}$ for which perturbative calculations do not make sense. Thus we are led to consider 
larger compactification
radii with $V_6 \sim 10^6 \ell_s^6$ which give $g^2 \sim 0.06$. One cannot have a smaller gauge coupling because in that case the blow up radius becomes larger than the compactification
radius. This value of the gauge coupling is quite interesting. For smaller couplings, which as we saw are hard to obtain, we can neglect the supergravity corrections in the potential in 
eq. (12). We also get a very  flat spectrum of density fluctuations, $n=1$. For larger couplings the supergravity corrections in the scalar potential are important and the spectrum of
density fluctuations are not necessarily very flat, $n \sim 0.98$. Moreover, in this case, the F--terms lead to a running spectral index with $n<1$ ($n>1$) at short (long) wavelengths[\FTE].
Clearly, our estimates which give us the borderline value for $g$ are not enough to decide which of these possibilities occur. For this, a detailed examination of cosmic 
string production in P--term models on D--branes is required.

A possible way to avoid the constraints in eq. (29) coming from cosmic strings is to have more than one complex scalar with nonzero VEV at the end of inflation[\ACH]. In this case the cosmic
strings that form are semi--local, i.e. they are not topological since the vacuum manifold is $S^3$. The number density of these strings after inflation vanishes (for equal gauge and 
Yukawa couplings) and therefore the constraint in eq. (29) does not apply. (For a more detailed examination of this isssue see [\DHK, \COM].)   
In our model, a second complex scalar with a VEV ($\Phi_-$) already exists when $f \not=0$, i.e. when there is an F--term in addition to the D--term. Even with only a
D--term this can be accomplished if there are two or more hypermultiplets in the model. The number of hypermultiplets in the bifundamental representaion is given by the strings which
connect D5 branes wrapped on different (and intersecting) $P^1$'s. For example with
three D5's on a $Z_3$ singularity (two wrapping one $P^1$ and the other wrapping the other second one) we get two pairs of complex scalars and therefore there will be no stable cosmic strings

From the bulk point of view, P--term inflation on the branes corresponds to the motion of the branes relative to each other, i.e. D--brane inflation. Consider two D5 branes along the 
$X_1,X_2,X_3,X_6,X_7$
directions on a $Z_3$ $ALE \times T^2$ (along the $X_6,X_7,X_8,X_9$ and $X_4,X_5$ directions respectively). Clearly, the two $P^1$'s that are blown up are along the $X_6,X_7$
directions. The complex deformation in eq. (16) describes the complex structure on the $P^1$'s which can be seen as the compactified $X_6,X_7$ plane. The branes can move along the
$X_4,X_5$ and $X_8,X_9$ directions. The motion along the former (latter) are described by the world--volume fields $S$ ($\Phi_{1,2})$. In other words, the values of $\Phi_{1,2}$ and
$S$ parametrize the Higgs and Coulomb branches respectively . However, the resolution and deformation of the singularity break supersymmetry and reduce the moduli space to a point 
(the supersymmetric
final state of P--term inflation). This supersymmetry breaking means that the two D5 branes feel an attractive force and start moving towards each other. This motion in the bulk 
describes P--term inflation on the world--volume. The attractive bulk potential corresponds on the world--volume to the inflaton mass which arises from the one--loop 
corrections to the superpotential. We see that two D5 branes initially separated along the $X_4,X_5$ directions will start to approach each other leading to inflation. In the meantime 
the branes start to separate along $X_8,X_9$. At the end of inflation the branes are at the same $X_4,X_5$ coordintes and separated along $X_8,X_9$.

Our scenario for P--term inflation on D--branes wrapped on deformed and resolved singularities can be easily generalized to more complicated spaces. First note that the $Z_3$ ALE singularity
is the simplest possible on for P--term inflation. The world--volume theory on D5 branes wrapped on a $Z_2$ singularity does not have hypermultiplets since in this case there is only
one blown--up $P^1$ and hypermultiplets arise from pairs of intersecting $P^1$'s. However, we can consider any $A_n$ ($Z_{n+1}$) type singularity which is described by the hypersurface
$$f(x,y,z)=x^2+y^2+z^{n+1}=0 \eqno(31)$$
The deformed singularity is given by
$$f(x,y,z,t_i)=x^2+y^2+\Pi_{i=1}^{n+1}(z+t_i) \qquad \qquad \Sigma_{i=1}^{n+1} t_i=0 \eqno(32)$$
The $n$ deformations are parametrized by $t_i$ and described by the ``holomorphic volumes'' $\alpha_i$ as in eq. (17). The resolution of the singularity is described by the blow--up
of $n$ $P^1$'s which intersect each other, each with a volume $v_i$ as in eq. (19). If we wrap $N_i$ D5 branes on the $n$ different $P^1$'s we get the gauge group $\Pi_{i=1}^n U(N_i)$ with
hypermultiplets in the bifundamental $(N_i, \bar N_j)$ and $(N_j, \bar N_i)$ representations; i.e. a quiver theory. The superpotential and the D--terms are simple generalization of 
those in eq. (4). Clearly, any $U(1)$ subgroup with a pair of bifundamentals would be sufficient to realize P--term inflation as we described above.

\bigskip
\centerline{\bf 4. P--term Inflation in Intersecting Brane Models}
\medskip

P--term inflation can also be realized in Hanany--Witten models[\HAN]. Unfortunately these cannot be compactified and therefore serve only as a realization of our model
close to the orbifold singularity. For simplicity, we consider the minimal model in section 2 which as we saw in section 3 is described by two 
D5 branes wrapped on a resolved and deformed) $Z_3$ ALE singularity. 

In terms of intersecting branes, the smooth $Z_3$ ALE space is described by three parallel NS5 branes along the 
$X_1,X_2,X_3,X_4,X_5$ directions and at the same $X_7,X_8,X_9$ coordinates. The two D5 branes wrapped on the blown up $Z_3$ singularity 
correspond to two D4 branes along the $X_1,X_2,X_3,X_6$
directions and stretched between the three NS5 branes (i.e. one D4 brane between the first and the second NS5 branes and the other one between the second and the third NS5 branes). 
The above intersecting brane configuration is T--dual to the one in section 3 given in terms of two wrapped D5 branes.
Under a T--duality along the $X_7$ direction the D4 branes become D5 branes. The three parallel NS5 branes become three five--dimensional Kaluza--Klein monopoles which are
described by the three--center Taub--NUT space[\NUT]
$$ds^2=V^{-1}(dz-A_i d y_i)^2+Vdy_i dy_i \eqno(33)$$
where
$$V=1+\Sigma_{r=1}^{3}{{2 \ell_s} \over {|y_r-y_{r_i}|}} \qquad \qquad \partial_i V=\epsilon_{ijk} \partial _j A_k \eqno(34)$$
Near the singularity, one can drop the constant term in $V$ and the metric becomes that of the $Z_3$ ALE space. This shows the equivalence of the two T--dual
descriptions (up to issues related to compactification). This ALE space is not singular but smooth with the $Z_3$ orbifold singularity blown up. The blow--up radii of the spheres
correspond to the distances between the NS5 branes. The D5 branes of section 3 correspond to the D5 branes
obtained after T--duality since these stretch along the $X_6,X_7$ directions which correspond to the blown--up spheres.
This description is similar to the ones that appear in refs. [\KAL] and [\LAS]. However, note that in our case there are no D6 branes; 
the hypermultiplets arise from strings that connect the two D4 branes separated by an NS5 brane. The absence of the D6 brane is the reason why in this model the transverse space can 
be compactified (in the T--dual picture).

The triplet of P--terms that give rise to P--term inflation are obtained by moving the NS5 branes to different $X_7,X_8,X_9$ coordinates. Defining $\Delta X_i=X_{i1}-X_{i3}$ where $1$ 
and $3$ denote the first and third NS5 branes respectively, we can choose 
$\Delta X_7$ to correspond to the D--term. Then, $\Delta X_8, \Delta X_9$ correspond to the other two P--terms (which appear as linear terms in the superpotential). 
It can be shown that the P--terms are given by
$$\xi_{1,2,3}={\Delta X_{7,8,9} \over {2 \pi \ell_s^2 g_{YM} L}} \eqno(35)$$
where $L$ is the distance between the two NS5 branes along the $X_6$ direction and the gauge coupling is
$$g^2_{YM}=(2 \pi)^2 g_s{\ell_s \over L} \eqno(36)$$
We see that the three P--terms are completely equivalent and the transformation in eq. (11) corresponds to a simple rotation in the $X_7,X_8,X_9$ space.
When coupled to ${\cal N}=1$ supergravity this symmetry is broken which is parametrized by the parameter $f$ 
$$f={{(\Delta X_8)^2+(\Delta X_9)^2} \over {(\Delta X_8)^2+(\Delta X_9)^2+(\Delta X_7)^2}} \eqno(37)$$ 

The matter content of the above brane configuration is well--known. The world--volume theory on the D4 branes has ${\cal N}=2$ supersymmetry. The gauge group is $U(1) \times U(1)$. 
The neutral scalars in these vector multiplets describe the positions the two D4 branes along the $X_4,X_5$ directions. The charged hypermultiplet describes the positions of the D4 branes
along the $X_7,X_8,X_9$ directions and the Wilson line along the $X_6$ direction. As before the sum of the two $U(1)$'s gives the center of mass motion of the D4 branes and is not
interesting for our purposes. The difference between the two $U(1)$'s describes the relative position of the branes and is the $U(1)$ that is relevant for P--term inflation. The neutral and 
charged hypermultiplets are also the ones that correspond to this $U(1)$ and are given by linear combinations of the original ones (as in section 3).

As we mentioned above, this description can be easily generalized to the case of $N$ D5 branes wrapped on a resolved and deformed $Z_n$ ALE space. In terms of intersecting branes,
this corresponds to $N$ D4 branes stretched between $n$ parallel NS5 branes; $N_i$ D4 branes are stretched between the $i^{th}$ and the ${i+1}^{th}$ NS5 branes. The matter content is 
a quiver theory exactly as the 
one described at the end of section 3. In this case, any $U(1)$ subgroup with two hypermultiplets realizes P--term inflation.

\bigskip
\centerline{\bf 5. Conclusions and Discussion}
\medskip

We have shown that P--term inflation is realized on D5 branes which are wrapped on resolved and deformed $Z_3$ orbifold singularities. On the brane world-volume theory, the resolution 
and deformation correspond to an anomalous D--term in the potential and a linear term in the superpotential both of which are necessary for P--term inflation. In the limit of vanishing 
resolution or deformation we get F or D--term inflation as expected. The model can easily be generalized to any $Z_n$ (with $n>3$) singularity which results in a quiver theory. Our model 
is T--dual to an intersecting brane model in which two D4 branes stretch between three parallel NS5 branes. The stringy parameters of the model are constrained by
the magnitude of the density of perturbations and possible contributions to this from cosmic strings.

We found that the strongest constraint on the parameters of the model arises from the possible contribution of cosmic strings to the density perturbations. Whether such strings are
created at the end of inflation and their properties depend on the topology of the vacuum manifold. For example, in D--term inflation, the strings would be local and therefore stable.
They would contribute to the density perturbations and constrain the model. On the other hand, in P--term inflation, there are two complex fields and the vacuum manifold is $S^3$. As a result,
the cosmic string created are semi--local and do not contribute to the density perturbations. Due to the many interesting possibilities and their observational effects
cosmic string production at the end of D--brane inflation and its experiemental signatures merit further study.

For very small resolutions and/or deformations, e.g. $\xi \sim 10^{-60}M_P^2$ the above model can describe the current nonzero vacuum energy as quintessence. In fact, this type of hybrid 
quintessence[\QUI] was considered in[\KAL]. However, this requires unnaturally small blow--up radii and/or complex deformations.

%\bigskip

%\centerline{\bf Acknowledgements}

\vfill

\refout

\end
\bye